\documentclass[10pt,twocolumn,twoside]{IEEEtran}

\usepackage{bm}
\usepackage[final]{graphicx}

\newcommand{\CN}[1]{{\mathcal CN} \left(#1 \right)}
\newcommand{\E}[1]{E\left\{ #1 \right\}}
\newcommand{\Eu}[2]{{\rm E}_{#1}\left\{ #2 \right\}}
\renewcommand{\Pr}[1]{{\rm P}\left( #1 \right)}

\newcommand{\tr}[1]{{\rm Tr}\left\{#1 \right\}}
\newcommand{\Rhv}{\widehat{\bf R}}

\newcommand{\diag}[1]{\textnormal{diag}\left( #1 \right)}

\newcommand{\Av}{{\bf A}}
\newcommand{\av}{{\bf a}}
\newcommand{\cv}{{\bf c}}
\newcommand{\Ev}{{\bf E}}
\newcommand{\ev}{{\bf e}}

\newcommand{\Iv}{{\bf I}}

\newcommand{\Mv}{{\bf M}}
\newcommand{\nv}{{\bf n}}

\newcommand{\Rv}{{\bf R}}
\newcommand{\Rtv}{{\bf \tilde{R}}}

\newcommand{\sv}{{\bf s}}
\newcommand{\vv}{{\bf v}}
\newcommand{\Vv}{{\bf V}}
\newcommand{\wv}{{\bf w}}

\newcommand{\xv}{{\bf x}}
\newcommand{\Xv}{{\bf X}}
\newcommand{\Xhv}{\widehat{\bf X}}

\newcommand{\zv}{{\bf 0}}

\newcommand{\tv}{{\bm \theta}}
\newcommand{\thv}{\hat{\bm \theta}}
\newcommand{\Tv}{{\bm \Theta}}
\newcommand{\psiv}{{\bm \psi}}
\newcommand{\Psiv}{{\bm \Psi}}

\newcommand{\defi}{\stackrel{\triangle}{=}}

\newcommand{\sigman}{\sigma_n^2}

\newcommand{\RMDL}{{\rm RMDL}}
\newcommand{\GMDL}{{\rm GMDL}}

\newcommand{\bfa}[1]{{\bf #1}}

\newtheorem{Lemma}{Lemma}
\newtheorem{PLemma}{Proof of Lemma}

\renewcommand{\(}{\left(}
\renewcommand{\)}{\right)}

\title{Estimation of the Number of Sources in Unbalanced Arrays via Information Theoretic Criteria}

\author{
Eran Fishler$^\ddag$ and H. Vincent Poor$^{\dag}$
\thanks{
   $^\dag$H. V. Poor is with the Department of Electrical Engineering, Princeton University,
   Princeton, NJ 08544, USA, Tel: (201) 258-1816, Fax: (201) 258-1468,
   {\tt e-mail: poor@princeton.edu}
   }
\thanks{
   $^\ddag$E. Fishler was with the Department of Electrical Engineering, Princeton University,
   Princeton, NJ. He is now with the Stern School of Business,
   New York University, NY, NY. {\tt e-mail: ef485@stern.nyu.edu}.
   }
\thanks{$^*$ This research was supported by the U.S. Office of Naval
Research under Grant No. N00014-03-1-0102.} }

\begin{document}

\maketitle
\begin{abstract}

Estimating the number of sources impinging on an array of sensors
is a well known and well investigated problem. A common approach
for solving this problem is to use an information theoretic
criterion, such as Minimum Description Length (MDL) or the Akaike
Information Criterion (AIC). The MDL estimator is known to be a
consistent estimator, robust against deviations from the Gaussian
assumption, and non-robust against deviations from the point
source and/or temporally or spatially white additive noise
assumptions. Over the years several alternative estimation
algorithms have been proposed and tested. Usually, these
algorithms are shown, using computer simulations, to have improved
performance over the MDL estimator, and to be robust against
deviations from the assumed spatial model. Nevertheless, these
robust algorithms have high computational complexity, requiring
several multi-dimensional searches.

In this paper, motivated by real life problems, a systematic
approach toward the problem of robust estimation of the number of
sources using information theoretic criteria is taken. An MDL type
estimator that is robust against deviation from assumption of
equal noise level across the array is studied. The consistency of
this estimator, even when deviations from the equal noise level
assumption occur, is proven. A novel low-complexity implementation
method avoiding the need for multi-dimensional searches is
presented as well, making this estimator a favorable choice for
practical applications.

\end{abstract}

\section{Introduction}

\subsection{Motivation}

The problem of estimating the number of sources impinging on a
passive array of sensors has received a considerable amount of
attention during the last two decades. The first to address this
problem were Wax and Kailath, \cite{Wax:85}. In their seminal work
\cite{Wax:85} it is assumed that the additive noise process is a
spatially and temporally white Gaussian random process. Given this
assumption the number of sources can be deduced from the
multiplicity of the received signal correlation matrix's smallest
eigenvalue \cite{Kaveh:87,Zhao:86}. In order to avoid the use of
subjective thresholds required by multiple hypothesis testing
detectors \cite{Anderson:63}, Wax and Kailath suggested the use of
the Minimum Description Length (MDL) criterion for estimating the
number of sources. The MDL estimator can be interpreted as a test
for determining the multiplicity of the smallest eigenvalue
\cite{Zhao:86}.

Following \cite{Wax:85}, many other papers have addressed this
problem (see, among
others~\cite{Abramovich:01,Cho:94,Fishler:00a,Wong:90,Wu:95,Zoubir:99}).
These papers can be divided into two major groups: the first is
concerned with performance analysis of the MDL estimator
\cite{Fishler:02,Kaveh:87,Xu:95,Zhang:89}, while the second is
concerned with improvements on the MDL estimator.

Papers detailing improvements of the MDL estimator can be found
quite extensively:
\cite{Abramovich:01,Brcich:02,Cho:94,Radich:95,Wang:85,Wu:95} is
only a partial list of such works. In many of these works the MDL
approach is taken, and by exploiting some type of prior knowledge,
performance improvement is achieved
\cite{Abramovich:03,Cho:94,Zoubir:99}. One of the assumptions
usually made is that the additive noise process is a spatially
white process, and the robustness of the proposed methods against
deviations from this assumption is usually assessed via computer
simulations \cite{Wu:95}. In general it can be observed that
methods which use some kind of prior information are robust, while
methods which are based on the multiplicity of the smallest
eigenvalue are non-robust. The reason for these latter estimators'
lack of robustness is that, when a deviation from the assumed
model occurs, the multiplicity of the smallest eigenvalue equals
one \cite{Wax:85}. Thus, one can not infer the number of sources
from the multiplicity of the received signal correlation matrix's
smallest eigenvalue. On the other hand, methods that are based on
some prior knowledge, e.g., the array steering vectors, usually
have high computational complexity, requiring several
multi-dimensional numerical searches \cite{Wax:91}. Moreover,
these methods are not necessarily consistent when some deviations
from the assumed model occur, although they exhibit good
robustness properties in simulations.

Efficient and robust estimation of the number of sources is very
important in bio-medical applications (see, for example
\cite{Niijima:02}, and references therein). For example, in one
such application it is of interest to estimate the number of
neurons reacting to a short stimulus. This is done by placing a
very large array of sensors over a patient's head, and recording
the brain activity as received by these sensors. In these
bio-medical problems no {\em a priori} knowledge (e.g., knowledge
of steering vectors) exists. Moreover, since different sensors are
at slightly different distances from the patient's skin, the noise
levels at the outputs of the sensors vary considerably. Thus
bio-medical applications are an example of one important class of
problem in which the additive noise is not necessarily spatially
homogeneous.

Although robust estimators for the number of sources exist, these
estimators require some {\em a priori} knowledge which is often
not avilable, and their computational complexity is large, as
noted above. Thus, computationally efficient and robust estimators
for the number of sources are of considerable interest. These
estimators should not require prior knowledge and should be
consistent even when deviations from the assumed model occur. Such
estimators for specific types of deviations from the assumed model
are developed in this paper. In particular, we consider the
situation in which the sensor noise levels are spatially
inhomogeneous. It will be shown that while traditional methods for
estimating the number of sources tend to over-estimate the number
of sources under these circumstances, our proposed estimator does
not have this tendency.

\subsection{Problem Formulation}
\label{ss. Problem Formulation}

Consider an array of $p$ sensors and denote by $\xv(t)$ the
received, $p$-dimensional, signal vector at time instant $t$.
Denote by $q<p$ the number of signals impinging on the array. A
common model for the received signal vector is
\cite{Wax:91,Fishler:02}:
   \begin{eqnarray}
      \xv(t) = \Av\sv(t) + \nv(t)\quad,\quad t=1,2,\ldots,N
   \end{eqnarray}
where $\Av = [\av(\psiv_1), \av(\psiv_2), \cdots, \av(\psiv_q)]$
is a $p\times q$ matrix composed of $q$ $p$-dimensional vectors,
and $\av(\psiv)$ lies on the array manifold $\{{\mathcal
A}=\bfa{a}(\psiv)|\psiv \in \Psiv \}$, where $\Psiv$ denotes a set
of parameters describing the array response. $\av(\psiv)$ is
called the array response vector or the steering vector and $\Av$
is referred to as the steering matrix, and $\psiv_i$ is a vector
of unknown parameters associated with the $i$th source.
$\sv(t)=[s_1(t)\;\cdots\;s_q(t)]^{T}$ is a white complex,
stationary Gaussian random processes, with zero means and positive
definite correlation matrix, $\Rv_\sv$; $\nv(t)$ is a temporally
white complex Gaussian vector random process, independent of the
signals, with zero mean and correlation matrix given by
$\diag{[\sigma_1^2,\sigma_2^2,\ldots,\sigma_p^2]}$, where
$\diag{[\sigma_1^2,\sigma_2^2,\ldots,\sigma_p^2]}$ denotes a
diagonal matrix with the vector
$[\sigma_1^2,\sigma_2^2,\ldots,\sigma_p^2]$ on its diagonal. This
correlation matrix represents the scenario in which each sensor
potentially faces a different noise level. Define
$\sigma^2=\frac{1}{p}\sum_{i=1}^p\sigma_i^2$,
$w_i=\sigma_i^2-\sigma^2$, and $\wv=[w_1,w_2,\ldots,w_p]$. The
additive noise correlation matrix can be described with the aid of
$\sigma^2$ and $\wv$ as follows
$\E{\nv(t)\nv^H(t)}=\sigma^2\bfa{I}+\diag{\wv}$. This alternate
representation simplifies some of the proofs and derivations in
the sequel. Note, that the vector $\wv$ represents a deviation
from the assumption that the noise level is constant across the
array. Finally, all the elements of the steering matrix,
$\bfa{A}$, are assumed to be unknown \cite{Wax:85}, with the only
restriction being that $\bfa{A}$ is of full rank. In the sequel
the Gaussian assumption will be eased.

We denote by $\tv_q$ the set of unknown parameters assuming $q$
sources, that is $\tv_q =
[\Rv_{\sv,q},\Av_q,\sigma^2_{n,q},\wv_q]$, where $\Rv_{\sv,q}$ is
the transmitted signals' correlation matrix assuming $q$ sources;
$\Av_q$ is the steering matrix assuming $q$ sources;
$\sigma^2_{n,q}$ is the white noise level; and $\wv_q$ is the
vector containing the parameters representing the deviations from
the spatially white noise assumption. The parameter space of the
unknown parameters given $q$ sources is denoted by $\Tv_q$. The
problem is to estimate $q$ based on $N$ independent snapshots of
the array output, $\xv_1=\xv(t_1), \ldots, \xv_N=\xv(t_N)$
\cite{Wax:85}.

\subsection{Information Theoretic Criteria and MDL Estimators}

An Information Theoretic Criterion (ITC) is an estimation
criterion for choosing between several competing parametric
models~\cite{Zhao:86}. Given a parameterized family of probability
densities, $f_{\Xv}\( \Xv | \tv_q \) ,\; \tv_q \in \Tv_q$ for
$\Xv=[ \xv_1, \ldots, \xv_N]$ and for various $q$, an ITC
estimator selects $\hat{q}$ such that \cite{Zhao:86}:
\begin{eqnarray}
   \hat{q}_{\rm ITC} = \arg\min_q {\rm ITC}(q) =
   \arg \min_q \left\{ -L\( \thv_q \) + {\rm penalty}(q) \right\}
   \label{e. basic MDL}
\end{eqnarray}
where $L\( \tv_q \)\defi \log f_{\Xv}\( \Xv | \tv_q \)$ is the
log-likelihood of the measurements, $\thv_q = \arg \max_{ \tv_q
\in \Tv_q } f_{\Xv}\( \Xv | \tv_q \)$ is the maximum likelihood
(ML) estimate of the unknown parameters given the $q$th family of
distributions, and ${\rm penalty}(q)$ is some general penalty
function associated with the particular ITC used. The MDL and AIC
estimators are given by ${\rm penalty}(q) = 0.5 |\Tv_q|\log\( N
\)$ and ${\rm penalty}(p) = |\Tv_q|$ respectively, where $| \Tv_q
|$ is the number of free parameters in $\Tv_q$
\cite{Akaike:73,Rissanen:78,Wax:85}. It is well known that,
asymptotically and under certain regularity conditions, the MDL
estimator minimizes the description length (measured in bits) of
both the measurements, $\Xv$, and the model,
$\thv_q$~\cite{Rissanen:84}, while the AIC criterion minimizes the
Kullback-Liebler divergence between the various models and the
true one. In the rest of the paper we will consider only the MDL
criterion, although other penalty functions can also be treated
similarly.

Although in many problems associated with array processing, e.g.,
direction of arrival (DOA) estimation, one has some prior
knowledge about the array structure, when estimating the number of
sources this prior knowledge is usually ignored
\cite{Wax:85,Wax:91}. The reason for this is that by ignoring the
array structure and assuming Gaussian signals and noise, and $\wv
\equiv \bf 0$, the resulting MDL estimator (\ref{e. basic MDL}),
termed here the Gaussian-MDL (GMDL) estimator \cite{Fishler:02},
has a simple closed form expression given by \cite{Wax:85}
   \begin{eqnarray}
      \hat{q}|_{\GMDL} = \arg \min_{q=0, \ldots, p-1}
      \left[-N\log\frac{\prod_{i=q+1}^p l_i}{\(
      \frac{1}{p-q}\sum_{i=q+1}^p l_i\)^{p-q}} \right.\nonumber \\
      +\left.
      \frac{1}{2}(q(2p-q)+1)\log N \right]
      \label{e. GMDL}
   \end{eqnarray}
where $l_1 \ge l_2 \ge \cdots \ge l_p$ are the eigenvalues of the
empirical received signal's correlation matrix, $\Rhv =
\frac{1}{N}\sum \xv_i\xv_i^H$. It is well known that when $\wv
\equiv \bf 0$ the GMDL estimator is a consistent estimator of the
number of sources, while when $\wv\neq 0$, the GMDL estimator,
(\ref{e. GMDL}), is not consistent and in fact, as the number of
snapshots approaches infinity, the probability of error incurred
by the GMDL estimator approaches one \cite{Fishler:02}.

Denote by ${\bf \mathcal  R}_{q}$ the set of all positive
definite, rank $q$, Hermitian, $p\times p$ matrices, and by ${\bf
\mathcal W}$ the set of all zero mean $p$-length vectors. Given
the assumptions made in the problem formulation, the MDL estimator
for estimating the number of sources, denoted hereafter as the
Robust-MDL (RMDL) estimator, is given by,
   \begin{eqnarray}
      &&\hat{q}_{\RMDL} = \arg \min_{q=0, \ldots, p-1} \nonumber\\
      &&\left\{
      N\log\left| \hat{\Av}_q\hat{\Rv}_{\sv,q}\hat{\Av}^H_q +
      \hat{\sigma}^2_{n,q} \Iv +\diag{\hat{\wv}_q} \right| \right.\nonumber \\
      &&\left.+ \tr{ \(
      \hat{\Av}_q\hat{\Rv}_{\sv,q}\hat{\Av}^H_q +
      \hat{\sigma}^2_{n,q} \Iv +\diag{\hat{\wv}_q} \)^{-1}\hat{\Rv}} \right.\nonumber\\
      &&+\left.
      \frac{1}{2}(q(2p-q) + p)\log N \right\}
      \label{d. RMDL}
   \end{eqnarray}
where $\hat{\Av}_q, \hat{\Rv}_{\sv,q},
\hat{\sigma}^2_{n,q},\hat{\wv}_q$ are the ML estimates of the
unknown parameters assuming $q$ sources, that is
   \begin{eqnarray}
      &&\hat{\Av}_q, \hat{\Rv}_{\sv,q},
      \hat{\sigma}^2_{n,q},\hat{\wv}_q = \arg \max_{\Av_q\Rv_{\sv,q}\Av_q^H
      \in {\bf \mathcal R}_q,
      \sigma^2_{n,q}>0, \wv_q \in {\bf \mathcal W}}
      \nonumber\\
      &&\left[   - N\log\left| \Av_q\Rv_{\sv,q}\Av^H_q +
      \sigma^2_{n,q} \Iv +\diag{\wv_q} \right| \right. \nonumber \\
      &&\left.+ \tr{ \(
      \Av_q\Rv_{\sv,q}\Av^H_q +
      \sigma^2_{n,q} \Iv +\diag{\wv_q} \)^{-1}\hat{\Rv}} \right].
   \end{eqnarray}
Note that since $\Av_q\Rv_{\sv,q}\Av_q^H\in {\bf \mathcal R}_q$,
by using eigen-decomposition we can write $\Av_q\Rv_{\sv,q}\Av_q^H
= \sum_{i=1}^q \lambda_i \vv_i\vv_i^H$, where $\{\vv_i\}$ is an
orthonormal set of vectors. Hence, the vector of unknown
parameters assuming $q$ sources is also given by

\begin{equation}
   \tv_q = [\lambda_1, \ldots, \lambda_q,\vv_1^T, \ldots,
   \vv_q^T,\wv^T,\sigman].
   \label{d. theta_q}
\end{equation}

\subsection{Organization of the Paper}

The rest of this paper is organized as follows: In Section II we
discuss the indentifiabilty of the estimation problem and we prove
the consistency of the RMDL estimator. In Section III we describe
a low-complexity algorithm for approximating the RMDL estimator,
(\ref{d. RMDL}), and we discuss the properties of this algorithm.
In Section IV we present empirical results. In Section V some
concluding remarks are provided.

\section{Identifiability and Consistency of the RMDL Estimator}

\subsection{Identifiability}

Consider a parameterized family of probability density functions
(pdf's) $f_{\Xv}(\xv|\tv),\;\tv\in\Tv$. This family of densities
is said to be {\em identifiable} if for every $\tv\neq \tv'$, the
Kullback-Liebler divergence between $f_{\Xv}\( \xv|\tv \)$ and
$f_{\Xv}\( \xv|\tv' \)$ is greater than zero, that is $D\(
f_{\Xv}\( \xv|\tv \) || f_{\Xv}\( \xv | \tv' \) \)
>0$, where $D(f(\xv)||g(\xv))=\int f\log \frac{f}{g}$ is the
Kullback-Leibler divergence between $f(\xv)$ and $g(\xv)$
\cite{Cover:91}. This condition insures that there is a one-to-one
relationship between the parameter space and the statistical
properties of the measurements.

The problem discussed in Section \ref{ss. Problem Formulation} is
a model order selection problem \cite{Rissanen:84}. This problem
is unidentifiable if it is possible to find for some $k\neq l$ two
points in the parameter space, $\tv_k\in\Tv_k$ and $\tv_l\in\Tv_l$
such that $f\( \cdot | \tv_k \) = f\( \cdot | \tv_l\)$.
Unfortunately, we can, in fact, identify two such points leading
to the conclusion that the estimation problem discussed in Section
\ref{ss. Problem Formulation} is unidentifiable. The received
signal's pdf is fully characterized by the received signal's
correlation matrix. Thus, in order to prove that the problem is
unidentifiable, it suffices to find two different parameter values
under which the corresponding received signal's correlation
matrices are equal. Take, for example, the following received
signal correlation matrix: $\diag{[11, 10.5, 9.5, 10]}$. This
correlation matrix can result from a noise-only scenario with
$\sigman = 10.25$ and $\wv = [0.75,0.25,-~0.75,-~0.25]$, or from a
one source scenario where $\sigman = 10, \wv = [0,0.5,-~0.5,0],
\av = [1,0,0,0]^T$, and $\Rv_{\sv} = 1$. Thus we have found two
scenarios, the first corresponding to a noise only scenario, and
the other corresponding to a one source scenario, such that the
distribution of the received signal vector is the same. Thus, this
example shows that the estimation problem formulated is
unidentifiable.

In order to make the estimation problem identifiable, all the
points having the same received signal pdf must be removed from
the parameter space except one. As is the custom in model order
selection problems, among all the points having the same received
signal pdf, the one with the smallest number of sources, that is,
the point with the lowest number of unknown parameters, is left in
the parameter space, and the remaining ones are deleted. The main
question that arises is whether most of the points in the
parameter space are identifiable or not. Fortunately, the answer
to this question is yes; that is, most of the points in the
parameter space are identifiable. The following lemma
characterizes all the unidentifiable points in the parameter
space.
   \begin{Lemma}
      \label{l. identifiability}
      Suppose $q<p$. Then $\tv_q$ is an unidentifiable point in the parameter
      space if and only if the matrix $\sum_{i=1}^{q}\vv_i\vv_i^H$ contains
      $\alpha \ev_j = [ \zv_{j-1}, 1 , \zv_{p-j}]$ as its $j$th row for some
      $j\in[1,\ldots,p]$,  where $\vv_i$ defined in
      (\ref{d. theta_q}).
   \end{Lemma}
   \begin{PLemma}
      See Appendix \ref{a. proof identfiability}
   \end{PLemma}

The proof of Lemma \ref{l. identifiability} provides an
interesting physical interpretation of the unidentifiable points.
In particular, it can be seen from the proof of the lemma that all
the unidentifiable points are similar to the above example used to
show that the problem is unidentifiable. That is, an
unidentifiable point corresponds to a scenario where there are,
say $q$ sources, and one of them is received at only one of the
sensors. Since this source can not be distinguished from a
deviation, from some nominal value, of the noise level in the
corresponding element, this scenario could be confused with a
different scenario having one fewer source, and an increase in the
noise level at the proper element. From a practical viewpoint,
this type of situation is a rarity.

\subsection{Consistency of the RMDL Estimator}

In the previous subsection it was proved that the estimation
problem defined in Section \ref{ss. Problem Formulation} is
unidentifiable. Nevertheless, it was also argued that only a small
portion of the points in the parameter space are unidentifiable,
meaning that by excluding these points from the parameter space
the problem becomes identifiable. For the rest of this paper, we
consider these points to be excluded from the parameter space.
Once the estimation problem has been shown to be identifiable, it
is possible to infer the number of sources from the measurements.
However for a specific estimator, the issue of consistency must be
considered.

In model order selection, the common performance measure is the
probability of error, that is $P_e = \Pr{\hat{q} \neq q}$
\cite{Zhao:86}. In what follows the RMDL estimator, (\ref{d.
RMDL}), is proven to be a consistent estimator, that is
$\lim_{N\rightarrow\infty} P_e = 0$.
   \begin{Lemma}
      \label{l. consistency}
      The RMDL estimator, (\ref{d. RMDL}), is a consistent estimator of the number
      of sources.
   \end{Lemma}
   \begin{PLemma}
      See Appendix \ref{a. proof consistency}
   \end{PLemma}

Deviations from the assumption of spatial homogeneity are part of
our general model. Thus, even if the noise levels at various
sensor are not equal, according to Lemma \ref{l. consistency} the
RMDL estimator, (\ref{d. RMDL}), is still consistent. That is, the
probability of error of the RMDL estimator still converges to zero
even in the presence of deviations from assumption of equal noise
levels.

It is well known that the GMDL estimator, (\ref{e. GMDL}), is a
non-robust estimator when the noise levels at the various sensor
are not equal, i.e., the probability of error of the GMDL
estimator approaches one as $N\rightarrow\infty$. Nevertheless, it
is known that the GMDL estimator is robust against statistical
mismodeling. Under very weak regularity conditions, if the
transmitted signal and/or the additive noise are non-Gaussian,
then the probability of error of the GMDL estimator still
converges to zero. Fortunately, it can be shown that the RMDL
estimator, (\ref{d. RMDL}) is robust against statistical
mismodeling as well. Being robust against both statistical and
spatial mismodeling is an advantage of the RMDL estimator over the
GMDL estimator.

We conclude this subsection by proving that the RMDL estimator,
(\ref{d. RMDL}) is a consistent estimator even in the presence of
statistical mismodeling. Denote by $g(\xv)$ the actual pdf of the
received signal at some time instant, and by $f\( \xv | \tv \)$
the \underline{assumed} measurement pdf, i.e., the Gaussian
distribution. Note that it is still assumed that $\Rv_\xv =
\E{\xv\xv^H}$ has the following form $\Rv_\xv =
\Av_\sv\Rv_\sv\Av^H + \sigman\Iv + \diag{\wv}$. Let
$\Eu{g}{h(\xv)} = \int h(\xv)g(\xv){\rm d}\xv$. The following
lemma establishes the consistency of the RMDL estimator when the
sources are not Gaussian
   \begin{Lemma}
      \label{l. robust consistency}
      Assume that $\Eu{g}{\frac{ \partial \log f\( \xv | \tv \)}
      { \partial \tv } \frac{\partial \log f\( \xv | \tv \)}
      {\partial\tv}^T }$ and $\Eu{g}{ \frac{ \partial^2 \log
      f\( \xv | \tv \)}{\partial\tv(\partial\tv)^T} }$ exist and are
      finite. Then the probability of error of the
      RMDL estimator converges to zero as $N\rightarrow\infty$.
   \end{Lemma}
   \begin{PLemma}
      See Appendix \ref{a. proof robust consistency}.
   \end{PLemma}

\section{A Practical Estimation Algorithm}

In the previous section the asymptotic properties of the RMDL
estimators were considered. It was proven that the RMDL estimator
is both a consistent and robust estimator of the number of
sources. These two properties make the RMDL estimator very
appealing for use in practical problems. However the computational
complexity of the RMDL estimator is still very high compared to
that of the GMDL estimator. Recall that in order to implement the
RMDL estimator ML estimates of the unknown parameters must be
found for every possible number of sources. Since no closed-form
expression for these ML estimates exists, multi-dimensional
numerical searches must be used in order to find them. Even for
moderate array sizes, e.g., $p=6$, the number of unknown
parameters is a few dozen, which makes the task of finding the ML
estimates impractical.

In order to overcome the computational burden of computing the ML
estimates, we propose to replace the ML estimates by estimates
obtained using a low-complexity estimation algorithm. A reasonable
criterion used in array processing applications is to choose as an
estimate the parameter vector that minimizes the Frobenius norm of
the error matrix \cite{Bohme:88,Wong:92}; that is
   \begin{eqnarray}
      \hat{\tv}_{q,LS} &=& \arg\min_{\tv_q\in\Tv_q} || (\Rhv -
      \Rv_\xv(\tv_q))||_F^2  \nonumber\\
      &=& \arg\min_{\tv_q\in\Tv_q} \tr{(\Rhv -
      \Rv_\xv(\tv_q))(\Rhv - \Rv_\xv(\tv_q))^H} \nonumber \\
      &=&
      \arg\min_{\tv_q\in\Tv_q} \sum_{i=1}^p\sum_{j=1}^p \left| [(\Rhv -
      \Rv_\xv(\tv_q))(\Rhv - \Rv_\xv(\tv_q)) ]_{ij} \right|^2
      \label{e. LS estimator}
   \end{eqnarray}
and the corresponding estimate for the number of sources is given
by,
   \begin{eqnarray}
      \hat{q} = \arg \min_{q=0,\ldots,p-1} \left [ -L\( \thv_{q,LS} \) +
      q(2p-q)\frac{\log N}{2} \right].
      \label{e. RMDL iterative}
   \end{eqnarray}

Replacing the ML estimates with their LS counterparts raises two
important questions. One is whether replacing the ML estimates
with the LS estimates results in performance loss; and the second
is whether efficient algorithms for computing the LS estimates
exist. Fortunately, it can be demonstrated that no performance
loss is incurred (asymptotically) by replacing the ML estimates
with the LS estimates, and an efficient algorithm for computing
the LS estimates exists. It was pointed out by one of the
reviewers that for finite sample sizes since the ML estimates are
replaced by the LS estimates, it is not guaranteed that $L\(
\thv_{q,LS} \)<L\( \thv_{q+1,LS} \)$. This problem can be easily
solved by noting that because the problem is a nested hypotheses
problem $hat{\theta}_q\in \Theta_{q+1}$. Therefore, if
$L(\hat{\theta}_{q|LS})< L(\hat{\theta}_{q+1|LS})$, we can use
$L(\hat{\theta}_{q|LS})$ instead of $L(\hat{\theta}_{q+1|LS})$ in
the MDL formula.

Our problem is a model order selection problem, and our main
interest is in the probability of error of the proposed estimator.
In \cite{Fishler:02}, it is demonstrated that the MDL's asymptotic
probability of error depends on $\tv_q$ and $\tv_{q-1}^\ast \in
\Tv_{q-1}$, where $\tv_{q-1}^\ast = \arg
\min_{\tv_{q-1}}D(f(\xv|\tv_q)||f(\xv|\tv_{q-1}))$. It is easily
seen that $\tv_{q-1}^\ast$ is the limit of the ML estimates under
the assumption of $q-1$ sources, i.e., $\hat{\tv_{q-1}}|\tv_q
\stackrel{N\rightarrow\infty}{\longrightarrow} \tv_{q-1}^\ast$.
Analysis similar to that in \cite{Fishler:02} demonstrates that if
a consistent estimator is used instead of the ML estimator in the
MDL estimator, then the asymptotic probability of detection
remains the same. Since the LS estimator is a consistent estimator
of the unknown parameters, the asymptotic performance of the
RMDL's simplified version, (\ref{e. RMDL iterative}), is the same
as the asymptotic performance of the RMDL estimator, (\ref{d.
RMDL}).

Similarly to the ML estimates, $\hat{\tv}_{q,LS}$ is the solution
of a nonlinear programming problem, requiring brute-force
multi-dimensional search. Nevertheless, based on the concept of
serial interference cancellation (SIC) \cite{Buzzi:01}, in what
follows a novel algorithm for finding $\hat{\tv}_{q,LS}$ is
suggested. In this algorithm the unknown parameters are divided
into two groups, and given the estimate of one group of unknown
parameters, an estimate of a second group of unknown parameters is
constructed. The estimates are constructed in such a way as to
insure a decrease in the Frobenius norm of the error matrix after
each iteration. The estimation process iterates between the two
groups of unknown parameters, until the estimates converge to a
stationary point.

In multiple access communications two (or more) users transmit
information over two non-orthogonal subspaces. The serial
interference cancellation multiuser detection algorithm for data
detection in such situations works as follows. First, the unknown
parameters associated with the first user are estimated. Next, an
error signal is constructed by subtracting from the received
signal the estimated first user's transmitted signal. In the next
stage, the unknown parameters associated with the second user are
estimated from the error signal. In the next iteration, the
unknown parameters associated with the first user are re-estimated
based on the received signal after subtraction of the estimated
second user's transmitted signal. This iterative process is
continued until convergence is reached.

The principle behind the SIC multiuser detector can be used for
constructing a novel low-complexity estimation algorithm for
estimating the unknown parameters in the present situation. In
what follows such a low-complexity estimation algorithm is
described and its properties are discussed. The unknown parameters
in our estimation problem are $[\lambda_1, \ldots,
\lambda_q,\vv_1^T, \ldots, \vv_q^T,\wv^T,\sigman]$, or
equivalently, $\Av_q\Rv_{\sv,q}\Av_q^H,\sigma_{n,q}^2,\wv_q$.
These unknown parameters are divided into two groups. The first
group contains $[\lambda_1, \ldots, \lambda_q,\vv_1^T, \ldots,
\vv_q^T,\sigman]$, or equivalently $\Av_q\Rv_{\sv,q}\Av_q^H$ and
$\sigma_{n,q}^2$, while the second contains $\wv_q$. The first
group corresponds to the unknown parameters of the ideal point
source plus spatially white additive noise model, while the second
corresponds to the unknown parameters representing the deviations
from the ideal model. In a sense, $\wv_q$ can be regarded as the
unknown parameters that robustify  the estimator. The input to the
algorithm is $\Rhv_\xv =
\frac{1}{N}\sum_{t=1}^N\xv(t_i)\xv(t_i)^H$ which is a sufficient
statistic for estimating the unknown parameters, assuming Gaussian
sources and noise.

Denote by $\sigma_{n,q}^{j},\Av_q^j\Rv_{\sv,q}^j\(\Av_q^j\)^H$,
and $\wv_q^j$ the estimates of the unknown parameters after the
$j$th iteration. The proposed algorithm is implemented as follows:
In the first iteration,
$\sigma_{n,q}^{1},\Av_q^1\Rv_{\sv,q}^1\(\Av_q^1\)^H$ are estimated
from $\Rhv_\xv$. The best estimates, in both the ML sense and the
Least Squares (LS) sense (see appendix \ref{a. convergence
algorithm}), are
   \begin{eqnarray}
      &&\sigma_{n,q}^1 = \sqrt{ \frac{1}{p-q} \sum_{i=q+1}^p l_i }
      \\
      &&\Av_q^1\Rv_{\sv,q}^1\(\Av_q^1\)^H = \sum_{i=1}^q \(l_i -
      \(\sigma_{n,q}^1\)^2 \)\vv_i\vv_i^H
   \end{eqnarray}
where $l_1 > \cdots >l_p$ and $\vv_1, \ldots,\vv_p$ are,
respectively, the eigenvalues and eigenvectors of $\Rhv_\xv$. In
the next step an error matrix, denoted by $\Ev$, is constructed by
subtracting from $\Rhv_\xv$ the estimate for the estimated part of
the received signal's correlation matrix corresponding to the
ideal model, that is $\Av_q^1\Rv_{\sv,q}^1\(\Av_q^1\)^H +
\(\sigma_{n,q}^1\)^2\Iv$. Thus, the error matrix is given by
  \begin{eqnarray}
      \Ev = \Rhv - \Av_q^1\Rv_{\sv,q}^1\(\Av_q^1\)^H -
      \( \sigma_{n,1}^1 \)^2\Iv.
   \end{eqnarray}
Next, $\wv_q^1$ is estimated from $\Ev$, and the best estimate in
both the ML and LS sense is,
   \begin{eqnarray}
      \wv_q^1 = \diag{\Ev}.
   \end{eqnarray}
At the $j$th iteration, we apply the same procedure except that
$\sigma_{n,q}^{j}$ and $\Av_q^j\Rv_{\sv,q}^j\(\Av_q^j\)^H$ are
estimated from $\Rhv- \wv_q^{j-1}$, while $\wv_q^j$ is estimated
from $\Rhv - \Av_q^j\Rv_{\sv,q}^j\(\Av_q^j\)^H - \( \sigma_{n,q}^j
\)^2 \Iv$.

Summarizing the above, our proposed estimation algorithm is given
as follows:
   \begin{enumerate}
      \item Initialize $\Ev = \Rhv$.
      \item Compute $l_1 \ge \cdots l_p$, $\vv_1, \ldots, \vv_p$
      the eigenvalues and the corresponding eigenvectors of
      $\Ev$.
      \item Compute the following estimates,
         \begin{eqnarray}
            &&\sigma_{n,q} = \sqrt{ \frac{1}{p-q} \sum_{i=q+1}^p l_i }
            \\
            &&\Av_q\Rv_{\sv,q}\(\Av_q\)^H = \sum_{i=1}^q \(l_i -
            \(\sigma_{n,q}\)^2 \)\vv_i\vv_i^H
            \\
            &&\wv_q = \diag{ \Rhv -
            \Av_q\Rv_{\sv,q}\(\Av_q\)^H - \(\sigma_{n,q}\)^2\Iv}
         \end{eqnarray}
      \item Compute $\Ev = \Rhv - \wv_q$.
      \item If the estimates have stabilized, stop; otherwise return to step 2.
   \end{enumerate}

A major question that arises is whether this algorithm is
guaranteed to converge and, if so, whether the stationary point of
the algorithm is optimal in some sense. Fortunately, the answers
to these questions are yes. In Appendix \ref{a. convergence
algorithm} it is proven that in each step of the algorithm, the
Frobenius norm of the error matrix decreases, that is $||\Rhv -
\Rv(\tv_q^n) ||_F^2 \ge ||\Rhv - \Rv(\tv_q^{n+1}) ||_F^2$, where
$\tv_q^n$ is the estimate of the unknown parameters after the
$n$th iteration. This also proves that the proposed algorithm
converge to a local minimum of the LS cost function.

Consider our proposed iterative algorithm. The most complex
operation in our algorithm is the eigenvalue decomposition whose
complexity is ${\mathcal O}(p^3)$. Since the process is repeated
$p$ times (one for each possible number of sources), the
complexity of our algorithm is ${\mathcal O}(p^4)$ per iteration.

Since no closed expression for the ML estimates exists, some
numerical maximization method must be used. Therefore, the
complexity of the ML estimator depends on the number of iterations
and the exact numerical maximization method used. However, we can
still demonstrate that the complexity of the ML estimator is
higher than that of our proposed algorithm. Since efficient
numerical maximization algorithms require the computation of the
derivative of the likelihood function, we examine the complexity
of computing this derivative. The most complex operation in
computing the derivative is $\frac{\partial
\tr{\Rv_{\xv}^{-1}(\tv)\Rhv}}{\partial \theta_i} =
-\tr{\left[\Rv_{\xv}^{-1}(\tv)\Rhv\Rv_{\xv}^{-1}(\tv)\right]\frac{\partial
\Rv_{\xv}^{-1}(\tv)}{\partial \theta_i}}$, which has a complexity
of ${\mathcal O}(p^8)$. This operation has to be repeated $p$
times, one for each possible number of sources. Therefore the
complexity of computing the derivative of the likelihood function
per iteration is ${\mathcal O}(p^9)$. It follows that for $p>3$,
the complexity of the ML estimator is higher by several orders of
magnitude than our proposed iterative algorithm.

\section{Simulations}

In this subsection simulation results with synthetic data are
presented. We consider a uniform linear array with 10 elements,
and assume three equal-power and independent sources having
signal-to-noise ratio (SNR) per element of $0\;dB$. The sources'
directions of arrival (DOA's) are taken to be $[0^\circ\;
5.7^\circ \; 11.4^\circ]$. We consider two cases: the first
corresponds to complex Gaussian sources, i.e., $s(t)\sim
\CN{0,\sigma^2}$; and the second corresponds to sources that are
distributed as complex Laplacian sources, i.e., $\Re \(s(t)\)$ and
$\Im\(s(t)\)$ are independent random variables having pdf
$\frac{1}{\alpha}e^{-\frac{|x|}{\alpha}}$. The second case
corresponds to impulsive sources usually found in bio-medical
application.

\begin{figure}[htbp]
      \centering
      \includegraphics[scale=0.45]{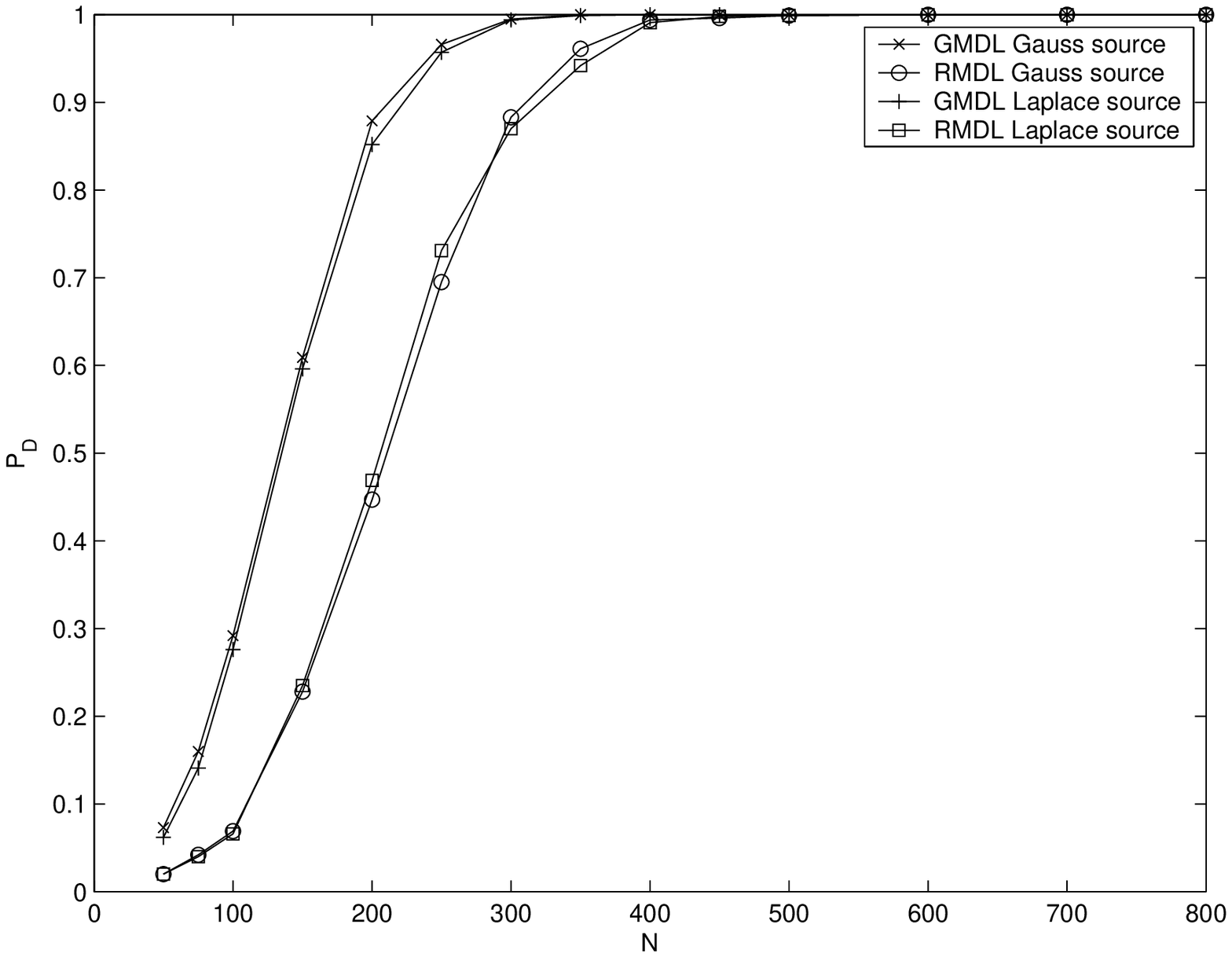}
      \caption{Three-user scenario, no mismatch. Probability of correct decision as a
      function of the number of snapshots.}
      \label{Fig:Graph1}
   \end{figure}

We first consider the case in which $\wv=0$; i.e., the noise is
spatially white. Figure \ref{Fig:Graph1} depicts the probability
of correct decision in this case of both the GMDL estimator and
the RMDL estimator when used with the estimates computed by the
iterative algorithm. Since no deviations from the spatial white
noise model exist in this case, the GMDL estimator is both
consistent and robust, and indeed the empirical probability of
error of the GMDL estimator converges to zero whether the sources
are Gaussian or Laplacian. The RMDL estimator is also both a
consistent and a robust estimator, and again the empirical
probability of error of the RMDL estimator converges to zero as
well, independent of the source distribution. These empirical
results demonstrate that the GMDL estimator is superior to the
RMDL estimator in this situation, an additional 100 samples are
required by the RMDL estimator in order to achieve the same
probability of correct decision as the GMDL estimator. In
\cite{Fishler:02a} it was proven that by exploiting more prior
information the performance of the MDL estimator improves. This
explains the superiority of the GMDL estimator over the RMDL
estimator, since the GMDL estimator makes use of the spatial
whiteness of the additive noise process, while the RMDL estimator
ignores this information.

\begin{figure}[htbp]
      \centering
      \includegraphics[scale=0.45]{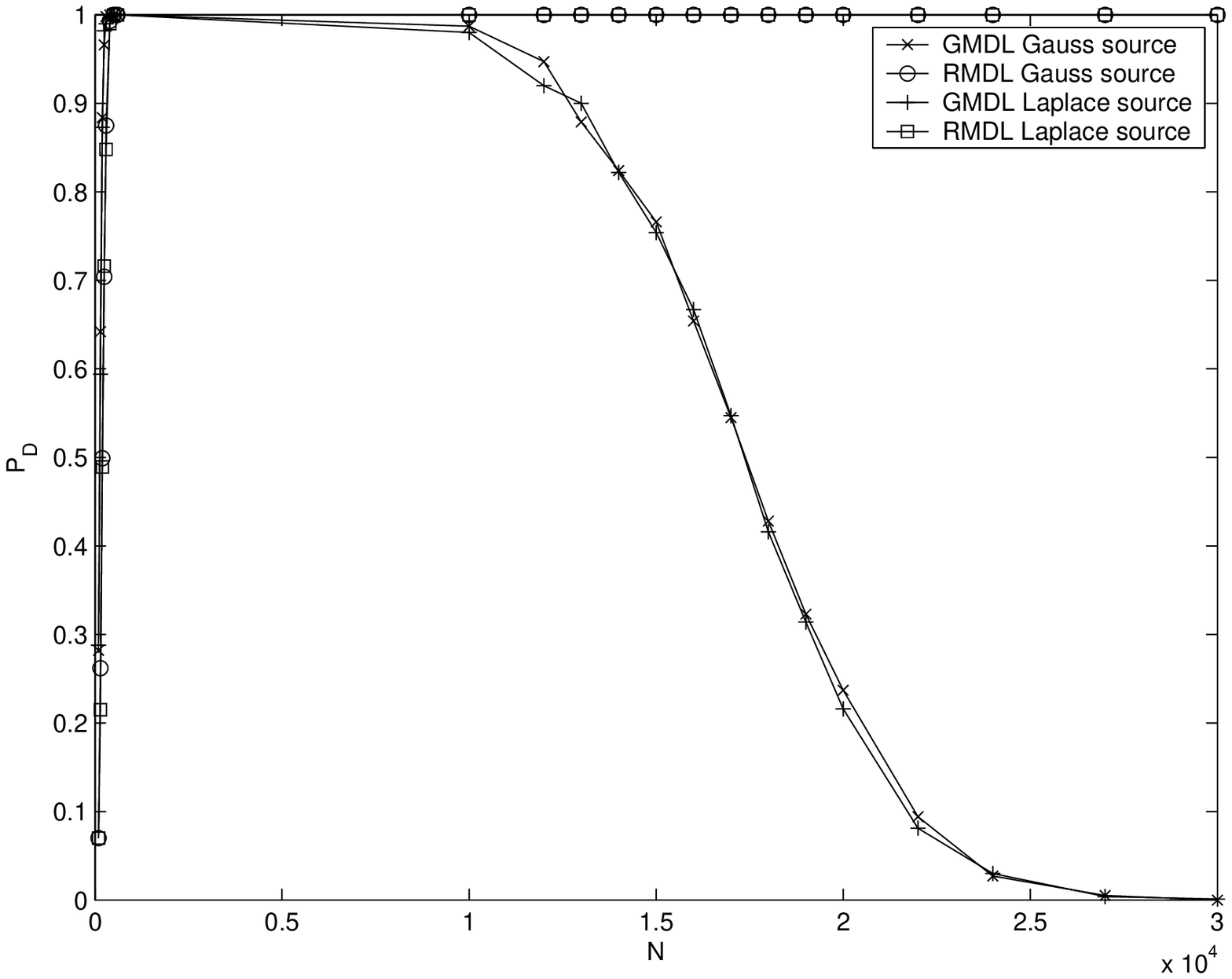}
      \caption{Three-user scenario, weak mismatch. Probability of correct decision as a
      function of the number of snapshots.}
      \label{Fig:Graph2}
   \end{figure}

In practice, multi-channel receivers are used in DOA estimation
systems. The noise level in each receiver is different and hence
the system has to be calibrated. Due to finite integration time,
errors and different drifts in each channel, small differences in
the noise levels at the different receiver channels exist. In the
next example this scenario is simulated. For simulating this
scenario $\wv$ is taken to be $\wv = \frac{\sigman}{10}[-9/10,
-7/10, \ldots, 9/30]$. This $\wv$ represents a scenario in which
the noise level in each receiver is different from the nominal
noise level by no more than $-10\,dB$. Figure \ref{Fig:Graph2}
depicts the probability of correct decision of both the GMDL and
the RMDL estimators as functions of the number of snapshots taken
for both Gaussian and Laplacian sources.

The multiplicity of the received signal correlation matrix's
smallest eigenvalue is equal to one, and hence the GMDL estimator
is not consistent, that is $\Pr{\hat{q} \neq 3} \rightarrow 1$
\cite{Zhao:86}. From Fig. \ref{Fig:Graph2} it is seen that the
empirical probability of error of the GMDL estimator converges to
one as the number of snapshots increases. Nevertheless, it can be
seen that this happen only when the number of snapshots is quite
large (about 10,000). This phenomenon can be explained by
examining the eigenvalues of the received signal's correlation
matrix. The eigenvalues of the received signal's correlation
matrix are given by $[20.1,10.9,1.93,1.07,\cdots,0.92]$. For the
GMDL estimator, the simulated scenario corresponds to a scenario
where $p-1$ sources exists, the noise level equals to $0.9$, and
the SNR of the fourth strongest source at the array output is
$-7\,dB$. The GMDL requires about 10,000 snapshots in order for
the probability of detection of this weak ``virtual" source to be
noticeable. As the number of snapshots increases, the probability
of detection of this weak virtual source increases as well,
causing the probability of correct decision to decrease to zero.
On the other hand, it can be seen that the probability of error of
the RMDL estimator converges to zero as the number of snapshots
increases for both the Gaussian and the Laplacian sources. This
demonstrate both the consistency and the robustness of the RMDL
estimator.

\begin{figure}[htbp]
      \centering
      \includegraphics[scale=0.45]{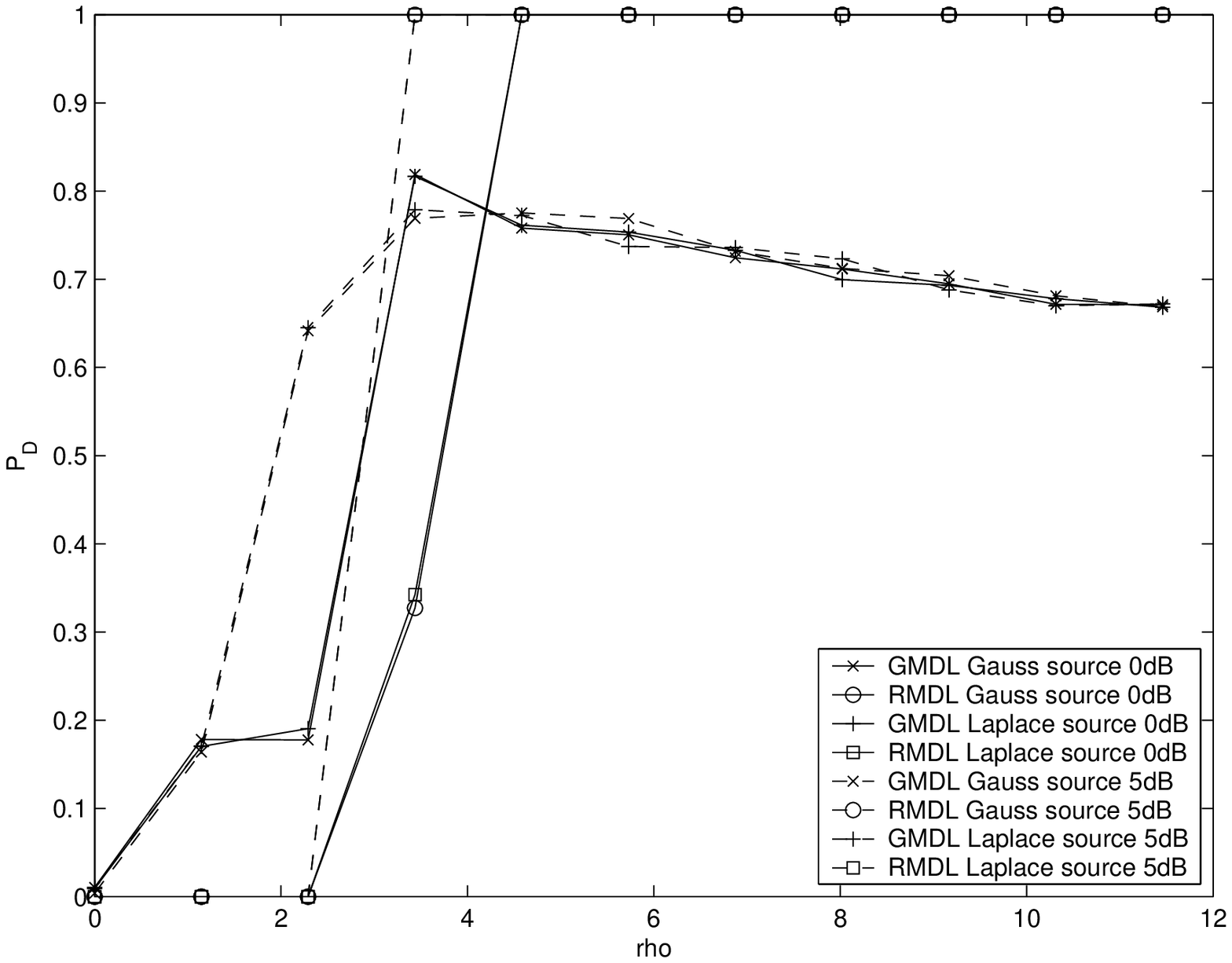}
      \caption{Three-user scenario, weak mismatch. Probability of correct decision as a
      function of the spatial separation between the sources.}
      \label{Fig:Graph2a}
   \end{figure}

In Figure \ref{Fig:Graph2a} we study the spatial separation
between the sources required for reliable detection. We assume
that the three sources' directions of arrival are
$[0,\rho,2\rho]$, 15,000 snapshots are taken by the receiver, and
the SNR per element is either $0\,dB$ or $5\,dB$. Figure
\ref{Fig:Graph2a} depicts the probability of correct decision of
both the GMDL and the RMDL estimators for both Gaussian and
Laplacian sources as a function of $\rho$.

In the figure we can see again that the RMDL estimator outperforms
the GMDL estimator. Even if large separation between the sources
exists, the probability of correct decision of the GMDL estimator
does not approach one. The probability of correct decision of the
RMDL estimator, on the other hand, approaches one with the
increase in the separation between the sources. This difference
can be explained with the aid of the received signal correlation
matrix's eigenvalue spectrum. The received signal correlation
matrix eigenvalues equal $[11.54,11.05,10.39,1.07,...,0.9237]$.
The three highest eigenvalues correspond to the three sources.
However, due to the different noise level in each sensor, the rest
of the eigenvalues are not equal to the noise level. The large
number of snapshots enables the GMDL estimator to detect the
differences in the weakest eigenvalues as valid sources, which
results in an error event. However, if the number of snapshots is
reduced, the GMDL estimator will not detect these differences.
Nonetheless, if the number of snapshot is reduced, and a valid
weak source exists, the GMDL estimator will not detect this valid
source.

\begin{figure}[htbp]
      \centering
      \includegraphics[scale=0.45]{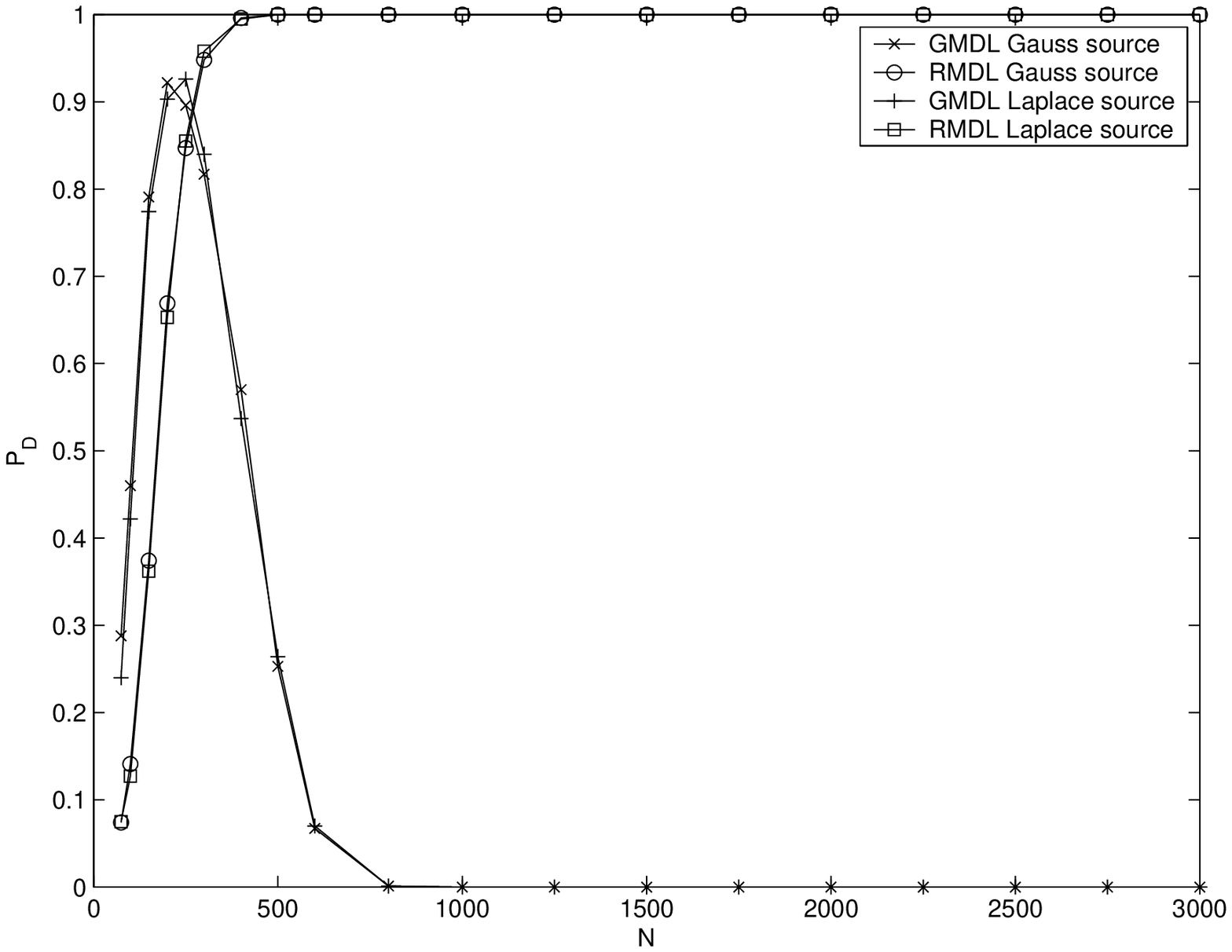}
      \caption{Three-user scenario, strong mismatch. Probability of correct decision as a
      function of the number of snapshots.}
      \label{Fig:Graph3}
   \end{figure}

As discussed in the beginning of this paper, in biological
applications the noise level may vary considerably between the
different receiver channels. Thus, large deviations from the ideal
model are expected in such systems. For simulating this type of
scenario we take $\wv = \frac{\sigman}{2}[-9/10, -7/10, \ldots,
9/10]$, which represents deviations of up to $-3\,dB$ from the
nominal noise level. Figure \ref{Fig:Graph3} depicts the
probabilities of correct decision of the GMDL and the RMDL
estimators as functions of the number of snapshots taken.

It can be seen that in this scenario the empirical error
probability of the GMDL estimator approaches  one even when the
number of snapshots is small (about 750). Again, this can be
explained by examining  the received signal correlation matrix's
eigenvalues, which are equal to $[20.13, 10.93, 2, 1.36,
\ldots,0.62]$. The GMDL estimator interprets this scenario as a
$p-1$ sources scenario with the noise level equal to $0.5$, and
the SNR of the fourth strongest source at the array output is
$6\,dB$. Due to its high SNR, only a small number of snapshots are
required for detecting this ``virtual" source, and by detecting
this virtual source an error event is created. As the number of
snapshots increases, the probability of detection of this virtual
source increases as well, causing the probability of correct
decision to decrease to zero. Again, it can be seen that the
probability of error of the RMDL estimator converges to zero as
the number of snapshots increases.

\begin{figure}[htbp]
      \centering
      \includegraphics[scale=0.45]{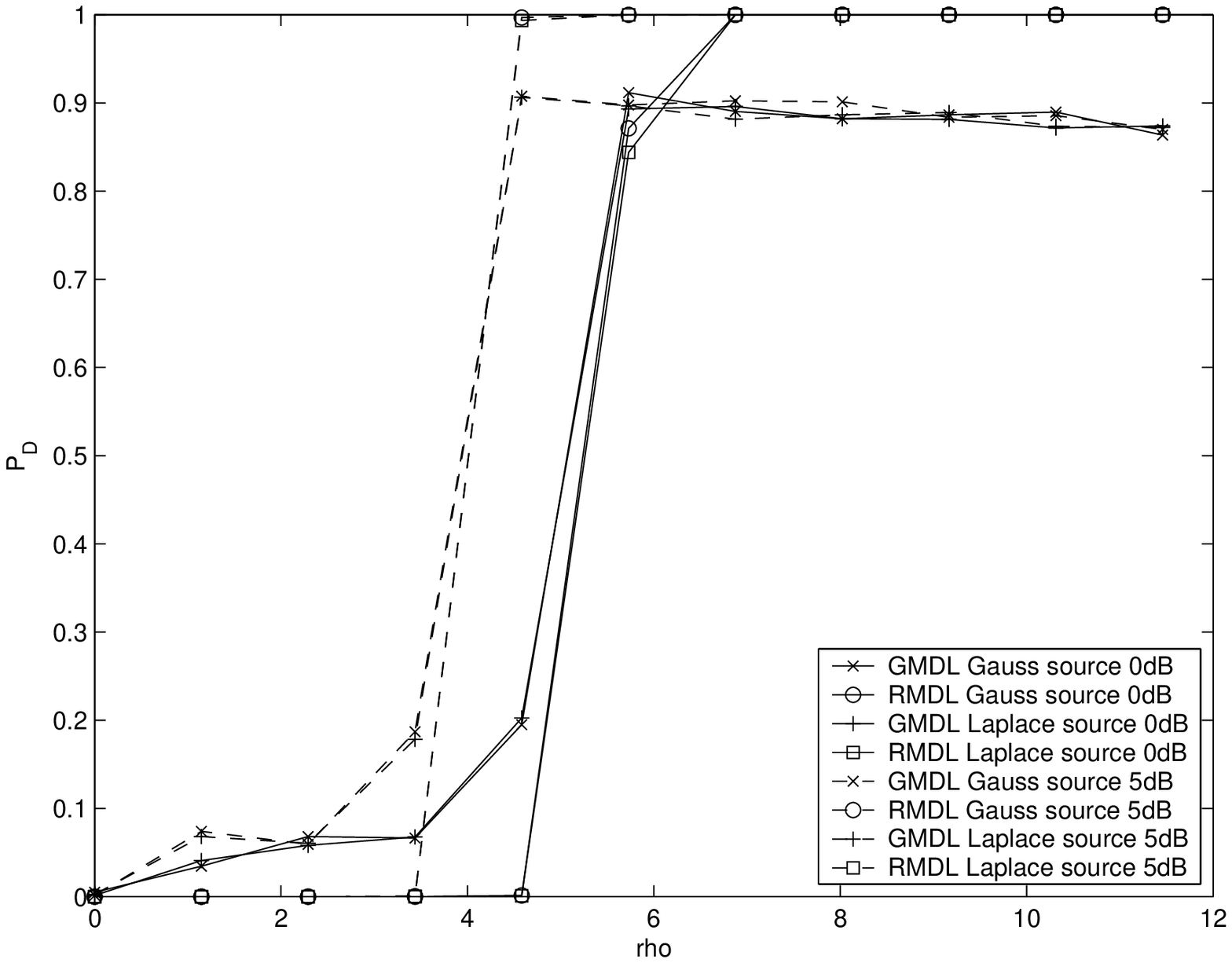}
      \caption{Three-user scenario, strong mismatch. Probability of correct decision as a
      function of the spatial separation between the sources}
      \label{Fig:Graph3a}
   \end{figure}

In the last figure, Figure \ref{Fig:Graph3a}, we study the spatial
separation between the sources required for reliable detection
when the deviation from the equal noise power assumption is large.
We assume that three sources' directions of arrival are
$[0,\rho,2\rho]$, 250 snapshots are taken by the receiver, and the
SNR per element is either $0\,dB$ or $5\,dB$. Figure
\ref{Fig:Graph3a} depicts the probabilities of correct decision of
both the GMDL and the RMDL estimators for the Gaussian and
Laplacian sources as a function of $\rho$. Again, we can see that
the RMDL estimator outperforms the GMDL estimator. Even for large
separation between the sources, the deviation from the equal noise
level assumption results in a change in the eigenvalue structure.
This change is detected by the GMDL estimator as an additional
sources, and hence an error event occurs.

\section{Summary and Concluding Remarks}

In this paper the problem of robust estimation of the number of
sources impinging on an array of sensors has been addressed. It
has been demonstrated that by proper use of additional unknown
parameters, the resulting estimator, denoted as the RMDL
estimator, is robust against both spatial and statistical
mismodeling. This situation represents an improvement on the
traditional MDL estimator which is robust only against statistical
mismodeling. In addition, a novel low-complexity algorithm for
computing the estimates of the unknown parameters has been
presented. It has been shown that this algorithm converges to the
LS estimates of the unknown parameters. On one hand, the
computational complexity of the proposed estimator is higher than
the complexity of the traditional MDL estimator; on the other hand
the complexity is far less than the complexity of known robust
estimators which require several multi-dimensional searches.

The proposed estimation algorithm can be used to robustify other
estimation algorithms as well. Take for example the MUSIC
algorithm for estimating DOAs \cite{Stoica:89}. It is well known
that the MUSIC algorithm is not robust against spatial
mismodeling. Even slight spatial mismodeling can cause a large
error in the estimated signal subspace, leading to substantial
estimation errors. The use of our estimation technique to improve
the robustness of the MUSIC algorithm is an interesting topic for
further study.

\bibliographystyle{ieeetr}
\bibliography{asp}

\appendices

\section{Proof of Lemma \ref{l. identifiability}}
\label{a. proof identfiability}

In this appendix Lemma \ref{l. identifiability} is proven by a way
of induction on the number of sources. We first note that since
$\sum_{i=1}^q\vv_i\vv_i^H$ is an Hermitian matrix, then if
$\sum_{i=1}^q\vv_i\vv_i^H$ contains $\ev_j$ as its $j$th row it
also contains $\ev_j^T$ as its $j$th column.

We first assume $q=0$; that is, the noise-only scenario. Since the
noise-only scenario is always identifiable, the lemma holds for
this case.

Now assume that the lemma holds for $q$ sources, that is for every
identifiable point $\tv_q\in\Tv_q$, $\sum_{i=1}^q \vv_i\vv_i^H$
does not have $\ev_j$ as one of its rows for every $j=1,\ldots,p$.

The following two lemmas will be essential in what follows.
   \begin{Lemma}
      Assume that, $\tv_q\in\Tv_q$ is an identifiable point, and
      denote by $\Rtv(\tv_q)\defi\sum_{i=1}^{q}\lambda_i\vv_i\vv_i^H$.
      Denote by $l_1 \ge \cdots \ge l_{q+1} > 0 = \cdots = 0$
      and $\{ \cv_i \}$ are
      the eigenvalues and their corresponding eigenvectors of
      the matrix $\Rtv(\tv_q)+\ev_j\ev_j^H$.
      Assume that rank$(\Rtv(\tv_q)+\ev_j\ev_j^H) =
      q+1$, then,
      $\sum_{i=1}^{q+1}\cv_i\cv_i^H$ has $\ev_j$ as his $j$th row.
      \label{l. aux l1}
   \end{Lemma}
   \begin{PLemma}
      Assume with out loss of generality that $j=1$. Since
      $\{\cv_i\}$ is an ortho-normal basis,
      $\sum_{i=1}^p\cv_i\cv_i^H=\sum_{i=1}^{q+1}\cv_i\cv_i^H +
      \sum_{i=q+2}^{p}\cv_i\cv_i^H =\Iv$. According to the lemma
      we have to prove that $\sum_{i=1}^{q+1}\cv_i\cv_i^H$ has
      the following form,
         \begin{eqnarray}
            \sum_{i=1}^{q+1}\cv_i\cv_i^H = \left [
               \begin{array}{cc}
                 1 & \zv \\
                 \zv^T & \Mv \\
               \end{array}
            \right].
         \end{eqnarray}
      This will happen if and only if $\sum_{i=q+2}^{p}\cv_i\cv_i^H$
      has the following form
         \begin{eqnarray}
            \sum_{i=q+2}^{p}\cv_i\cv_i^H = \left [
               \begin{array}{cc}
                  0 & \zv \\
                  \zv^T & \Mv'
               \end{array}
            \right ],
         \label{app. tmp1}
         \end{eqnarray}
      where $\Mv+\Mv'=\Iv$, (recall that $\sum_{i=1}^{q+1}\cv_i\cv_i^H +
      \sum_{i=q+2}^{p}\cv_i\cv_i^H =\Iv$). It is easy to verify
      that $\sum_{i=q+2}^{p}\cv_i\cv_i^H$ will have the form given
      by (\ref{app. tmp1}) if and only if
      $[\cv_l]_1 = 0$ for every $l>q+1$, so proving the lemma is
      equivalent to proving that $[\cv_l]_1 = 0$ for every
      $l>q+1$.
      Assume that $l>q+1$. From the properties
      of eigen-decomposition it follows that
         \begin{eqnarray}
            (\Rtv(\tv_q)+\ev_1\ev_1^H)\cv_l = \sum_{i=1}^{q+1} l_i
            \cv_i\cv_i^H\cv_l = \zv = \nonumber\\
            \( \sum_{i=1}^q \lambda_i
            \vv_i\vv_i^H + \ev_1\ev_1^H\) \cv_l =
            \sum_{i=1}^q \lambda_i(\vv_i^H\cv_l)\vv_i +
            \ev_1[\cv_l]_1
            \label{e. lemma aux1 e1}
         \end{eqnarray}
      where $\lambda_1 \ge \cdots \ge \lambda_p$ and
      $\vv_1,\ldots,\vv_p$ are, respectively, the eigenvalues and
      eigenvectors of $\Rtv(\tv_q)$. Since $\{\cv_i\}_{i=1}^{q+1}$
      spans the subspace spanned by $\{\vv_i\}_{i=1}^q$, then
      $\vv_i^H\cv_l=0$ for every $i\le q$. Thus, by using
      (\ref{e. lemma aux1 e1}),
         \begin{eqnarray}
            \ev_1[\cv_l]_1 = \zv,
         \end{eqnarray}
      which is possible if and only if $[\cv_l]_1=0$. 
   \end{PLemma}
   \begin{Lemma}
      rank$(\Rtv(\tv_q)+\ev_i\ev_i^H) = q+1$.
      \label{l. aux l2}
   \end{Lemma}
   \begin{PLemma}
      Without loss of generality (wlg) it is proven that
      rank$(\Rtv(\tv_q)+\ev_1\ev_1^H) = q+1$. Assume that
      rank$(\Rtv(\tv_q)+\ev_1\ev_1^H) = q$. Thus the rank of both
      $\Rtv(\tv_q)$ and $\Rtv(\tv_q)+\ev_1\ev_1^H$ are equal.
      Hence, it is possible to find $k$ constants,
      denoted by $a_1, \ldots, a_q$, not all of them equal to zero,
      such that
         \begin{eqnarray}
            \ev_1 = \sum_{i=1}^q a_i\vv_i.
            \label{e. lemma aux2 e1}
         \end{eqnarray}
      From (\ref{e. lemma aux2 e1}) it is easy to see that
         \begin{eqnarray}
            \Rtv(\tv_q)+\ev_1\ev_1^H = \Vv\Av\Vv^H
         \end{eqnarray}
      where $\Vv=[\vv_1, \ldots, \vv_q]$, and $\Av$ is some
      $q\times q$ diagonal matrix. Since $\tv_q$ is an identifiable point,
      according the induction assumption
      there exists $l>q$ such that $[\vv_l]_1\neq 0$ (otherwise
      according to the previous lemma the point would have been
      unidentifiable contredicting our assumption that $\tv_q$ is
      identifiable). As such,
         \begin{eqnarray}
            \(\Rtv(\tv_q)+\ev_1\ev_1^H\)\vv_l = \Vv\Av\Vv^H\vv_l =
            \zv = \ev_1[\vv_l]_1
         \end{eqnarray}
      which is possible if and only if $[\vv_l]_1=0$,
      This is a contradiction, and Lemma \ref{l. aux l2} follows.
   \end{PLemma}

Define $g(\tv_q)$ to be a function taking as an argument an
identifiable point in $\Tv_q$, and returning a subset of
$\Tv_{q+1}$, such that for every $\tv_{q+1} \in g(\tv_q)$,
$\Rv_\xv(\tv_{q+1}) = \Rv_\xv(\tv_q)$, and for every $\tv_{q+1}
\in \overline{g(\tv_k)}$, $\Rv_\xv(\tv_{q+1}) \neq
\Rv_\xv(\tv_q)$. It is easy to see from Lemma \ref{l. aux l2} that
$\tv_{q+1} \in g(\tv_q)$ if and only if, $\Rtv(\tv_{q+1}) =
\Rtv(\tv_q) + [\wv(\tv_q)]_i\ev_i\ev_i^H, \sigman(\tv_{q+1}) =
\sigman(\tv_q)+\frac{1}{p-1}\sum_{j\neq i}[\wv(\tv_k)]_j$, and
$[\wv(\tv_{q+1})]_k = [\wv(\tv_q)]_k-\frac{1}{p-1}\sum_{j\neq
i}[\wv(\tv_q)]_j$ where $k\neq i$, and $1 \le i \le p$. From Lemma
\ref{l. aux l2} it is easy to see that $(\Rtv(\tv_{q+1}))$ has
rank $q+1$, and from Lemma \ref{l. aux l1} it is easy to see that
the conditions stated in Lemma \ref{l. identifiability} are
necessary. Since every unidentifiable point belongs to some
$g(\tv_q)$ then the lemma is proved. 

\section{Proof of Lemma \ref{l. consistency}}
\label{a. proof consistency}

In this appendix the consistency of the RMDL estimator is proved.
Specifically it is shown that the probability of error of the RMDL
estimator converges to zero as the number of snapshots increases
to infinity. An error event will occur if and only if there exists
$k\neq q$ such that $\RMDL(q) - \RMDL(k) > 0$. Thus in order to
prove the lemma it suffice to prove that for every $k\neq q$,
$\Pr{ \RMDL(q) - \RMDL(k)
> 0} \rightarrow 0$.

Assume that $k>q$. Since the problem is a nested hypothesis
problem, $\log f_{\Xv}(\Xv | \thv_k ) < \log f_{\Xv}(\Xv |
\thv_{p-1} )$ \cite{Fishler:02}. Also, since $\thv_q$ maximizes
the likelihood of the measurements under the assumption of $q$
sources, $\log f_{\Xv}(\Xv | \thv_q ) > \log f_{\Xv}(\Xv | \tv_q
)$, where $\tv_q$ is the true parameter value. Thus $\RMDL(q) -
\RMDL(k)$ can be bounded as follows,
   \begin{eqnarray*}
      \RMDL(q) - \RMDL(k) = \nonumber\\
      -\log f_{\Xv}(\Xv | \thv_q ) + \log
      f_{\Xv}(\Xv | \thv_k ) + \nonumber\\ (q(2p-q) - k(2p-k) )\frac{\log
      N}{2} \nonumber\\
      \le -\log f_{\Xv}(\Xv | \tv_q ) + \log
      f_{\Xv}(\Xv | \thv_{p-1} ) \nonumber\\
      + (q(2p-q) - k(2p-k) )\frac{\log
      N}{2}.
   \end{eqnarray*}

Using the spectral representation theorem, the received signal's
correlation matrix is equal to $\Rv_{\xv}(\tv_q) = \sum_{i=1}^p
\lambda_i \vv_i\vv_i^H$, where $\lambda_1 > \cdots > \lambda_p$
and $\vv_1, \ldots, \vv_q$ are, respectively, the eigenvalues and
their corresponding eigenvectors of $\Rv_{\xv}(\tv_q)$. Thus there
exists a point, denoted by $\tv_{p-1}^* \in \Tv_{p-1}$ such that
$\Rv_{\xv}(\tv_q) = \Rv_{\xv}(\tv_{p-1}^*)$ (take $A = [\vv_1,
\ldots, \vv_{p-1}],\Rv_{\sv} = \diag{\lambda_1 - \lambda_p,
\ldots, \lambda_2 - \lambda_1}, \sigman = \lambda_1, \wv = {\bf
0}$). Since $\tv_{p-1}^*$ is an inner point of $\Tv_{p-1}$, one
can use the theory of likelihood \cite{Lehmann:59} to show that
asymptotically, $-2\log f_{\Xv}(\Xv | \tv_q ) + 2\log f_{\Xv}(\Xv
| \thv_{p-1} ) = 2\log f_{\Xv}(\Xv | \thv_{p-1} ) -2\log
f_{\Xv}(\Xv | \tv_{p-1}^* )$ is distributed as a chi-square random
variable with degrees of freedoms equal to the number of unknown
parameters, $(p^2-1)$. We next note that since $q<k$, $(q(2p-q) -
k(2p-k) )\frac{\log N}{2} \rightarrow -\infty$ as $N$ approaches
to infinity. Thus, as the number of measurements increases, the
probability that $-\log f_{\Xv}(\Xv | \tv_q ) + \log f_{\Xv}(\Xv |
\thv_{p-1} )$ exceeds $|(q(2p-q) - k(2p-k) )\frac{\log N}{2}|$ is
given by the tail of the chi-square distribution, which approaches
zero as $N$ approaches to infinity. Thus,
   \begin{eqnarray}
      &&\Pr{\RMDL(q) - \RMDL(k) > 0 } < \nonumber \\
      &&{\rm Pr} \(-\log f_{\Xv}(\Xv | \tv_q ) + \log
      f_{\Xv}(\Xv | \thv_{p-1} ) \right.
      \nonumber\\
      &&+\left. (q(2p-q) - k(2p-k) )\frac{\log
      N}{2} > 0\) \stackrel{N\rightarrow\infty}{\rightarrow} 0,
      \label{e. a. t1}
   \end{eqnarray}
which complete the first part of the consistency proof.

Now, assume $k<q$. It was previously shown that under very weak
conditions the probability of miss of every MDL estimator
converges to zero as $N\rightarrow\infty$ \cite{Fishler:02}. In
particular the probability of miss of the RMDL estimator, which
satisfies the condition stated in \cite{Fishler:02} is the MDL
estimator, converges to zero as $N\rightarrow\infty$.

\section{Proof of Lemma \ref{l. robust consistency}}
\label{a. proof robust consistency}

The proof of Lemma \ref{l. robust consistency} is very similar to
the proof of Lemma \ref{l. consistency}, and thus only the
necessary modifications for the proof of lemma \ref{l.
consistency} are detailed. Again, in order to prove that the
probability of error converges to zero we will prove that ${\rm
Pr}\left\{\RMDL(q) - \RMDL(k) > 0\right\} \rightarrow 0$.

Assume $k>q$. It is easy to see from the proof of Lemma \ref{l.
consistency} that ${\rm Pr}\(\RMDL(q) - \RMDL(k) > 0 \) < {\rm
Pr}( -\log f_{\Xv}(\Xv | \tv_q ) + \log f_{\Xv}(\Xv | \thv_{p-1} )
+ (q(2p-q) - k(2p-k) )\frac{\log  N}{2} > 0)$. It is known that
asymptotically, given the conditions stated in the lemma, $\log
f_{\Xv}(\Xv | \tv_{p-1} ) - \log f_{\Xv}(\Xv | \thv_{p-1}^* )$ is
distributed as a weighted sum of chi-square random variables
having one degree of freedom \cite{Kent:82}. Thus by implying the
same reasoning used in the proof of Lemma \ref{l. consistency}, it
easily shown that $\Pr{\RMDL(q) - \RMDL(k)
> 0} \rightarrow 0$.

Assume $k<q$. Again, this case is a special case of a more general
theorem presented in \cite{Fishler:02} and hence we omit a
specific proof for this case.

\section{Convergence of the Proposed Estimation Algorithm}
\label{a. convergence algorithm}

Denote by $\thv_q^{n} =
[\hat{\wv}_{q,n},\hat{\sigma}^2_{q,n},\widehat{
\Av_{q,n}\Rv_{\sv,q,n}\Av_{q,n}^H}]$ the estimate of $\tv_q$ after
the $n$th iteration, and by $E_n$ the error between $\Rhv$ and
$\Rv_\xv(\thv_q^n)$, that is $E_n = \Rhv - \Rv_\xv(\thv_q^n)$. In
this appendix it is proven that $\tr{E_nE_n^H} >
\tr{E_{n+1}E_{n+1}^H}$. The following lemma will be very helpful
in the sequel.

   \begin{Lemma}
      Let $\Xv$ be a $p\times p$ Hermitian matrix, with eigenvalue
      representation $\Xv = \sum_{i=1}^p \alpha_i\vv_i\vv_i^H$.
      The closest (in the Frobenius norm sense)
      $p\times p$ Hermitian matrix $\Xhv$, such that $\Xhv =
      \sum_{i=1}^ql_i\cv_i\cv_i^H + l\sum_{i=q+1}^p\cv_i\cv_i^H$
      is the matrix $\widehat{\Xv} = \sum_{i=1}^q \alpha_i
      \vv_i\vv_i^H + \sum_{i=q+1}^p \frac{\sum_{j=q+1}^p l_j}{p-q} \vv_i\vv_i^H$
   \end{Lemma}
   \begin{PLemma}
      For the sake of simplicity we prove the lemma for real
      vectors, and not complex ones. The extension to complex
      vector is straight forward and thus is omitted here.
      We first note that we have to find the matrix $\Xhv$ such
      that $\tr{\(\Xv-\Xhv\)\(\Xv-\Xhv\)^T}$ is minimized. We
      note the following identities,
         \begin{eqnarray*}
            &&\Xv\Xv^T = \sum_{i=1}^p\sum_{j=1}^p \alpha_i\alpha_j
            \vv_i\vv_i^T\vv_j\vv_j^T =
            \sum_{i=1}^p\alpha_i^2\vv_i\vv_i^H \\
            &&\tr{\Xv\Xv^T} = \sum_{i=1}^p\alpha_i^2\\
            &&\Xhv\Xhv^T=\sum_{i=1}^{q+1}\sum_{j=1}^{q+1}
            l_il_j\cv_i\cv_i^T\cv_j\cv_j^T +
            \sum_{i=1}^{q+1}\sum_{j=q+1}^p l_i
            l\cv_i\cv_i^T\cv_j\cv_j^T \nonumber\\
            &&+
            \sum_{i=q+1}^p\sum_{j=1}^{q+1}l
            l_j\cv_i\cv_i^T\cv_j\cv_j^T +
            \sum_{i=q+1}^p\sum_{j=q+1}^p l^2
            \cv_i\cv_i^T\cv_j\cv_j^T \\
            &&\tr{\Xhv\Xhv^T} = \sum_{i=1}^{q+1}\sum_{j=1}^{q+1}
            l_il_j\(\cv_i^T\cv_j\)^2 +
            \sum_{i=1}^{q+1}\sum_{j=q+1}^p l_i
            l\(\cv_i^T\cv_j\)^2\nonumber\\
            &&+
            \sum_{i=q+1}^p\sum_{j=1}^{q+1}l
            l_j\(\cv_i^T\cv_j\)^2 +
            \sum_{i=q+1}^p\sum_{j=q+1}^p l^2
            \(\cv_i^T\cv_j\)^2 \\
            &&\Xv\Xhv^T =
            \sum_{i=1}^p\sum_{j=1}^q \alpha_i l_j \vv_i\vv_i^T
            \cv_j\cv_j^T + \sum_{i=1}^p\sum_{j=q+1}^p \alpha_i l
            \vv_i\vv_i^T \cv_j\cv_j^T \\
            &&\tr{\Xv\Xhv^T} =
            \sum_{i=1}^p\sum_{j=1}^q \alpha_i l_j \(\vv_i^T
            \cv_j\)^2 + \sum_{i=1}^p\sum_{j=q+1}^p \alpha_i l
            \(\vv_i^T \cv_j\)^2.
         \end{eqnarray*}
      By using these identities, $\tr{\(\Xv-\Xhv\)\(\Xv - \Xhv\)^H}$
      can be expressed as follows:
         \begin{eqnarray}
            &&R\defi\tr{\(\Xv-\Xhv\)\(\Xv \Xhv\)^T} \nonumber\\
            &&=\tr{\Xv\Xv^T} -
            2\tr{\Xv\Xhv^T} + \tr{\Xhv\Xhv^T} = \nonumber\\
            &&\sum_{i=1}^p\alpha_i^2
            -2\sum_{i=1}^p\sum_{j=1}^q \alpha_i l_j \(\vv_i^T
            \cv_j\)^2
            -2 \sum_{i=1}^p\sum_{j=q+1}^p \alpha_i l
            \(\vv_i^T \cv_j\)^2
            \nonumber\\
            &&+ \sum_{i=1}^{q+1}\sum_{j=1}^{q+1}
            l_il_j\(\cv_i^T\cv_j\)^2
            +\sum_{i=1}^{q+1}\sum_{j=q+1}^p l_i
            l\(\cv_i^T\cv_j\)^2 \nonumber\\
            &&+ \sum_{i=q+1}^p\sum_{j=1}^{q+1}l
            l_j\(\cv_i^T\cv_j\)^2
            + \sum_{i=q+1}^p\sum_{j=q+1}^p l^2
            \(\cv_i^T\cv_j\)^2.
         \end{eqnarray}
      The derivatives of $R$ with respect to the unknown parameters are given
      by the following:
         \begin{eqnarray*}
            &&\frac{\partial R}{\partial l_k} =
            -2\sum_{i=1}^p \alpha_i \(\vv_i^T \cv_k\)^2
            +2 l_k \(\cv_k^T\cv_k\)^2
            + \sum_{i\neq k}
            l_i\(\cv_i^T\cv_k\)^2
            \nonumber\\
            &&+\sum_{j=q+1}^p
            l\(\cv_k^T\cv_j\)^2
            + \sum_{i=q+1}^p l
            \(\cv_i^T\cv_k\)^2\,,\,k=1,\ldots,q \nonumber\\
            &&\frac{\partial R}{\partial l} =
            -2 \sum_{i=1}^p\sum_{j=q+1}^p \alpha_i
            \(\vv_i^T \cv_j\)^2
            +\sum_{i=1}^{q+1}\sum_{j=q+1}^p l_i
            \(\cv_i^T\cv_j\)^2
            \nonumber\\
            &&+ \sum_{i=q+1}^p\sum_{j=1}^{q+1}
            l_j\(\cv_i^T\cv_j\)^2
            + 2 \sum_{i=q+1}^p\sum_{j=q+1}^p l
            \(\cv_i^T\cv_j\)^2 \nonumber \\
            &&\frac{\partial R}{\partial \cv_k} =
            -4\sum_{i=1}^p \alpha_i l_k \(\vv_i^T
            \cv_k\)\vv_k
            + 4l_j^2\(\cv_k^H\cv_k\)\cv_k
            \nonumber\\
            &&+ 8\sum_{i\neq k}^{q+1}
            l_il_k\(\cv_i^T\cv_k\)\cv_i \,,\,k=1,\ldots q
            \nonumber\\
            &&\frac{\partial R}{\partial \cv_k} =
            -4 \sum_{i=1}^p \alpha_i l
            \(\vv_i^T \cv_k\)\cv_i
            +4\sum_{i=1}^{q+1} l_i
            l\(\cv_i^T\cv_k\)\cv_i
            \nonumber\\
            &&+4 l^2 \(\cv_k^T\cv_k\)\cv_k
            + 4\sum_{i=q+1,i\neq k}^p l^2
            \(\cv_i^T\cv_k\)\cv_i
         \end{eqnarray*}
      It is now easy to verify that by substituting into the above
      equations the proposed solution and exploiting the fact that $\{\vv_i\}$ is an
      orthonormal bases, all the derivatives are equal
      to zero, and hence the proposed solution minimizes $\tr{\(
      \Xv - \Xhv\)\(\Xv-\Xhv\)^H}$. 
   \end{PLemma}

According to the algorithm, at the beginning of the $(n+1)$th
iteration the following matrix is created, $\sum_{i=1}^q (l_i -
\sigman)\vv_i\vv_i^H + \sigman\Iv = \sum_{i=1}^ql_i\vv_i\vv_i^H +
\sigman\sum_{i=q+1}^p\vv_i\vv_i^H $, where $l_1> \cdots
> l_p$ and $\vv_1, \ldots, \vv_p$ are, respectively, the
eigenvalues and the corresponding eigenvectors of the matrix
   \begin{eqnarray}
      \Rhv - \diag{\hat{\wv}_{q,n}} = E_n +
      \widehat{\Av_{q,n}\Rv_{\sv,q,n}\Av_{q,n}^H}
      +\hat{\sigma}^2_{q,n}\Iv,
      \label{a. e. t41}
   \end{eqnarray}
and $\sigma_n^2 = \frac{1}{p-q}\sum_{i=q+1}^p l_i$. Denote by
$E_{n+1}'$ the error between $\Rhv - \diag{\hat{\wv}_{q,n}}$ and
$\sum_{i=1}^q (l_i - \sigman)\vv_i\vv_i^H + \sigman\Iv$; that is
$E_{n+1}' =\Rhv - \diag{\hat{\wv}_{q,n}} - \sum_{i=1}^q (l_i -
\sigman)\vv_i\vv_i^H - \sigman\Iv$. According to the Lemma 6
$\tr{E_{n+1}'E_{n+1}'^H} < \tr{E_n E_n^H}$.

At the second part of the $(n+1)$th iteration, $\wv_{q,n+1}$ is
constructed as follows,
   \begin{eqnarray}
      &&\wv_{q,n+1} = \diag{ \Rhv - \sum_{i=1}^q (l_i -
      \sigman)\vv_i\vv_i^H + \sigman\Iv} \nonumber\\
            &&= \diag{E_{n+1}' +
      \wv_{q,n}}.
      \label{a. e. t42}
   \end{eqnarray}
The total error, between $\Rhv$ and the estimate is
   \begin{eqnarray}
      &&E_{n+1} = \Rhv - \diag{\wv_{q,n}} - \sum_{i=1}^q (l_i -
      \sigman)\vv_i\vv_i^H \nonumber\\
            &&+ \sigman\Iv - \diag{\wv_{q,n+1}} =
      E_{n+1}' - \diag{E_{n+1}'}.
   \end{eqnarray}
Hence $\tr{E_{n+1}E_{n+1}^H} = \sum_{i,j}
[E_{n+1}]_{ij}[E_{n+1}]_{ij}^H = \sum_{i\neq j}
[E_{n+1}']_{ij}[E_{n+1}']_{ij}^H \le \sum_{i,j}
[E_{n+1}']_{ij}[E_{n+1}']_{ij}^H = \tr{ E_{n+1}'E_{n+1}'^H} \le
\tr{E_nE_n^H}$, which concludes the proof.

\end{document}